# Superconductivity and hybrid soft modes in TiSe$_2$

M. Maschek[1], S. Rosenkranz[2], R. Hott[1], R. Heid[1], D. Zocco[1], A. H. Said[3], A. Alatas[3], G. Karapetrov[4], Shan Zhu[5], Jasper van Wezel[5] and F. Weber[1]

[1]Karlsruhe Institute of Technology, Institute for Solid State Physics, 76131 Karlsruhe, Germany
[2]Materials Science Division, Argonne National Laboratory, Argonne, Illinois, 60439, USA
[3]Advanced Photon Source, Argonne National Laboratory, Argonne, Illinois, 60439, USA
[4] Department of Physics, Drexel University, 3141 Chestnut Street, Philadelphia, Pennsylvania 19104, USA
[5] Institute for Theoretical Physics, Institute of Physics, University of Amsterdam, 1090 GL Amsterdam, The Netherlands

**The competition between superconductivity and other ground states of solids is one of the challenging topics in condensed matter physics. Apart from high-temperature superconductors [1,2] this interplay also plays a central role in the layered transition-metal dichalcogenides, where superconductivity is stabilized by suppressing charge-density-wave order to zero temperature by intercalation [3] or applied pressure [4-7]. *1T*-TiSe$_2$ forms a prime example, featuring superconducting domes on intercalation as well as under applied pressure. Here, we present high energy-resolution inelastic x-ray scattering measurements of the CDW soft phonon mode in intercalated Cu$_x$TiSe$_2$ and pressurized *1T*-TiSe$_2$ along with detailed ab-initio calculations for the lattice dynamical properties and phonon-mediated superconductivity. We find that the intercalation-induced superconductivity can be explained by a solely phonon-mediated pairing mechanism, while this is not possible for the superconducting phase under pressure. We argue that a hybridization of phonon and exciton modes in the pairing mechanism is necessary to explain the full observed temperature-pressure-intercalation phase diagram. These results indicate that *1T*-TiSe$_2$ under pressure is close to the elusive state of the excitonic insulator.**

# I. Introduction

New superconducting materials are usually classified as conventional or unconventional superconductors when superconductivity can be explained via a phonon-mediated pairing mechanism or not, respectively. Whereas the former is present in elemental metals throughout the periodic table as well as in a plethora of more complex compounds, unconventional superconductivity is typically only found in the vicinity of a, most often magnetically, ordered phase. Specifically, suppressing this order via an extrinsic parameter such as pressure, chemical substitution or intercalation leads to an emergent superconducting dome centered on the critical point of the suppressed phase. Understanding the impact of the fluctuations in the vicinity of such a quantum critical point on the emergent superconducting phase is a paramount effort in solid state physics.

It is now evident that a number of layered materials exhibiting charge-density-wave (CDW) order, a periodic modulation of the charge carrier density and the atomic lattice, become superconducting when CDW order is suppressed.[3,4,8-10] Classic examples are the members of the transition-metal dichalcogenide family $MX_2$, where M = Nb, Ti, Ta, Mo and X = S, Se, which show a large diversity of CDW ordered phases competing with superconductivity [10,11]. *1T*-TiSe$_2$ is particularly interesting: Electronic correlations [12-14] as well as electron-phonon-coupling (EPC) [15,16] were reported to drive the CDW transition at $T_{CDW} \approx 200$ K. Moreover, *1T*-TiSe$_2$ features superconducting domes induced by pressure [4] as well by Cu-intercalation [3] and the discussion on the driving mechanism of the CDW directly extends to the nature of the superconducting phases, which could be mediated by excitons, phonons, or hybrid modes [17].

Recently it was reported that the phase diagram of *1T*-TiSe$_2$ under pressure is different from that of Cu-intercalation as well as from most unconventional superconductors. X-ray diffraction [5] revealed that the superconducting phase under pressure is completely surrounded by the CDW ordered phase rather than being centered on the critical pressure [4]. In this work we use inelastic x-ray scattering to investigate the superconducting pairing mechanisms in Cu$_x$TiSe$_2$ and *1T*-TiSe$_2$ under pressure. Employing *ab-initio* lattice dynamical calculations we show that the superconducting transition temperatures in Cu$_x$TiSe$_2$ are explained solely in terms of the EPC, but we demonstrate that this is not possible for superconductivity under pressure, which therefore requires an excitonic component in its pairing mechanism. A model theory explaining the qualitative features of the reported phase diagram is provided.

## II. Methods

### A. Sample synthesis

Samples were prepared via the iodine gas transport method[18] with copper intercalation during crystal growth. The copper contents were analyzed in energy dispersive x-ray spectroscopy measurements. Measured crystals had dimensions of $0.2 - 1$ mm in the *a-b* plane and $50 - 100$ µm along the crystallographic $c$ axis.

### B. Density functional perturbation theory

DFT calculations were performed in the framework of the mixed basis pseudopotential method.[19] The exchange-correlation functional was treated in the local-density approximation (LDA). Norm-conserving pseudo-potentials for Ti and Se were constructed with Ti $3s$ and $3p$ semicore states included in the valence space. Phonon energies and electronic contributions to the phonon line width were calculated using the linear response technique or density functional perturbation theory (DFPT)[20] in combination with the mixed-basis pseudopotential method.[21] To resolve fine features related to the Fermi surface geometry, Brillouin-zone (BZ) integrations were performed with a dense hexagonanal 24x24x8 **k**-point mesh (244 points in the irreducible BZ). The standard smearing technique was employed with a Gaussian broadening of $0.1$ eV $\leq \sigma \leq 0.5$ eV.

For $\sigma \geq 0.15$ eV, calculations were performed in the trigonal high temperature structure ($P\bar{3}m1$, #164) whereas for $\sigma < 0.15$ eV, calculations were performed in a $2 \times 2 \times 2$ supercell corresponding to the observed low temperature structure in the CDW ordered state. Detailed structural parameters are given in Appendix A.

### C. Inelastic x-ray scattering

The high-resolution inelastic x-ray scattering (IXS) experiments were carried out at the XOR 3-ID and 30-ID HERIX beamlines of the Advanced Photon Source, Argonne National Laboratory. The incident energies were 21.66 keV (3-ID) and 23.78 keV (30-ID). The horizontally scattered beam was analyzed by spherically curved silicon analysers with typical energy resolutions of 2.4 meV (3-ID) and 1.5 meV (30-ID) (full width at half maximum (FWHM)). The components ($Q_h$, $Q_k$, $Q_l$) of the scattering vector are expressed in reciprocal lattice units (r.l.u.) ($Q_h$, $Q_k$, $Q_l$) = (h*2π/a, k*2π/a, l*2π/c); with the lattice constants (for $x = 0$) $a = b = 3.54$ Å and $c = 6.01$ Å of the trigonal unit cell ($P\bar{3}m1$, #164).

Measured energy spectra were fitted using a pseudo-Voigt function for the elastic line with a variable amplitude and fixed line shape determined by scanning a piece of plastic and double-checked with scans through the CDW superlattice peak at base temperature in case of finite $T_{CDW}$. The phonon was fitted by a damped harmonic oscillator (DHO) function, where the energy $\omega_q$ of the damped phonon is given by $\omega_q = \sqrt{\Omega_q^2 - \Gamma_q^2}$ (Ref. [22]), where $\Omega_q$ is the unrenormalized phonon energy and $\Gamma_q$ denotes the phonon line width.

IXS measurements were conducted on $1T$-TiSe$_2$ samples loaded inside diamond anvil cells (DAC). The pressure was increased using a helium gas membrane system. The DAC was mounted to a Sumitomo cold head mounted on standard Huber Euler Cradle. The lowest temperature achieved with the DAC mounted is 5 K. The pressure was determined in-situ using a Ruby fluorescence system. The hydrostatic conditions were achieved using neon gas as the pressure transmitting medium

### III. DFPT

A natural question concerning the phase diagram is in what way does EPC, which leads to the CDW transition, contribute to superconductivity when the CDW order is suppressed. Theoretically, we address this question by ab-initio lattice dynamical calculations using density-functional-perturbation-theory (DFPT). This method is a routinely used tool to investigate EPC and related superconducting properties [23-30]. EPC is investigated based on the momentum and energy resolved electronic contribution to the phonon line width [24]

$$\gamma_{q,\vartheta} = 2\pi \omega_{q\vartheta} \sum_{k\nu\nu'} \left[ \left| g^{q\vartheta}_{k+q\nu',k\nu} \right|^2 \delta(\epsilon_{k\nu} - \epsilon_F) \cdot \delta\left(\epsilon_{k+q\nu'} - \epsilon_F\right) \right] \quad (1)$$

for a phonon mode $\vartheta$ at a given wave vector $q$ with energy $\omega_{q\vartheta}$ and EPC matrix element $g^{q\vartheta}_{k+q\nu',k\nu}$, in which $k$ and $\nu$ denote the wave vector and band index of the involved electronic states, respectively. Averaging over $\vartheta$ and $q$, we extract the Eliashberg function

$$\alpha^2 F(\omega) = \frac{1}{2\pi N(\epsilon_F)} \sum_{q,\vartheta} \frac{\gamma_{q,\vartheta}}{\omega_{q,\vartheta}} \delta(\omega - \omega_{q,\vartheta}), \quad (2)$$

the effective phonon frequency

$$\omega_{log} = \exp\left(\frac{2}{\lambda} \int_0^\infty d\omega \frac{\alpha^2 F(\omega)}{\omega} \ln \omega \right) \quad (3)$$

and the EPC constant

$$\lambda = 2 \int_0^\infty d\omega \frac{\alpha^2 F(\omega)}{\omega}, \quad (4)$$

where $N(\epsilon_F)$ is the density of states at the Fermi energy $\epsilon_F$. We can estimate the superconducting transition temperature $T_c$ by solving the full gap equations employing a typical value for the effective electron-electron interaction, e.g., $\mu^* = 0.1$.

In previous work [16] we have shown that DFPT successfully predicts the structural instability associated with the CDW transition in $1T$-TiSe$_2$ including the softening of a transversely polarized phonon mode, the so-called soft mode, at the L point, i.e. the zone boundary in the [111] direction of the trigonal Brillouin zone. Because of the finite momentum mesh used in the DFPT calculations, a numerical smearing $\sigma$ of the electronic bands is necessary for computational convergence. Qualitatively, it can be interpreted as thermal smearing of the electronic structure and has been used to simulate temperature dependent behavior [31-33]. Here, we notice that the CDW structural instability can be suppressed by increasing the smearing parameter $\sigma$ within our calculations. We therefore treat $\sigma$ as a numerical tuning parameter within the zero-temperature DFPT technique, which we employ to investigate the behavior of the superconducting transition temperature as the CDW transition is suppressed. The combination of these aspects makes DFPT a powerful tool to study the existence of a superconducting phase near the critical point of the CDW order.

Calculating the free energy for the trigonal high temperature structure as well as for the $2 \times 2 \times 2$ supercell corresponding to the structural distortion in the CDW phase, we find that the distorted phase of the CDW is stable for $\sigma < 0.15$ eV and energetically unfavorable for $\sigma \geq 0.15$. Relaxing the internal atomic coordinates within the supercell, the predicted structural distortion is in good agreement with recent theoretical and experimental work [33,34]. In the following, the lattice dynamical calculations are done using the $2 \times 2 \times 2$ supercell describing the CDW phase for small $\sigma$ values, and the high temperature trigonal structure for $\sigma \geq 0.15$ eV. In the supercell calculation (Fig. 1a), the $L$ points of the high temperature structure together with the CDW soft mode are backfolded to the zone center (ellipsoid in Fig. 1a). Hence, we find strong EPC in three phonon modes corresponding to the three different $L$ points of the high temperature structure. Because of the distorted atomic structure, the three modes are not degenerate anymore but distributed in energy, e.g. 6 meV $\leq E \leq$ 9 meV for $\sigma = 0.1$ eV (Fig. 1a). The EPC of these modes is still significant, as indicated by the large calculated linewidths $\gamma_q$ (vertical bars in Fig. 1b). Upon increasing $\sigma$ DFPT predicts a softening of all three modes on approaching the critical value of $\sigma \approx 0.15$ eV (Fig. 1b).

For $\sigma = 0.15$ eV we find a nearly soft phonon mode at the $L$ point (Fig. 1c) indicating the close vicinity to the critical point of the CDW phase transition line.[1] The soft mode quickly hardens going away from $\sigma = 0.15$ eV both on increasing and decreasing $\sigma$ as indicated by the dashed line in Figure 1b. DFPT predicts that only the soft mode is strongly affected by varying $\sigma$, along with its dispersion along the [001] direction, i.e. the $L - M$ line (Fig. 1c), which is a signature of the quasi two-dimensional structure of $1T$-TiSe$_2$.

---

[1] In Ref. [16], DFPT calculations shown for $\sigma = 0.15$ eV, showing an even lower soft mode energy of 0.25 meV, were actually performed using $\sigma = 0.148$ eV.

The calculated Eliashberg function $\alpha^2 F(\omega)$ and EPC constant $\lambda(\omega)$ show pronounced changes as function of $\sigma$ in the low energy range $\omega \leq 15$ meV (Fig. 2a-c). The hardening of the soft mode, on either side of the critical point, is accompanied by a removal of the dominant contribution to $\alpha^2 F(\omega)$, resulting in a strong suppression of the EPC constant. This demonstrates that in $1T$-TiSe$_2$ the CDW soft mode carries the largest part of the total EPC in contrast to the situation, e.g., in $2H$-NbSe$_2$ [35]. Solving the full gap equations, we estimate the superconducting transition temperature $T_c$ based on a scenario of phonon-mediated pairing. Only an estimate of $T_c$ can be given, because the value of the effective electron-electron interaction potential $\mu^*$ is not calculated within DFPT. For the estimated values shown in Figure 2d, we used $\mu^* = 0.1$, which was also used in previous calculations of $T_c$ in superconducting TiSe$_2$ [15] and $\mu^* = 0.15$ based on a recent experimental work on Cu$_x$TiSe$_2$ [36]. We find the maximum values of $T_c$ very close to the critical point of the CDW phase (Fig. 2d) irrespective of employing $\mu^* = 0.1$ ($T_{c,max} = 6.01$ K) or $\mu^* = 0.15$ ($T_{c,max} = 4.39$ K). Hence, DFPT demonstrates that in $1T$-TiSe$_2$ a phonon-mediated superconducting dome should appear when the CDW order is suppressed and this dome should be centered on the CDW critical point.

## IV. Cu$_x$TiSe$_2$

The most direct way of testing the scenario discussed above is by measuring the lattice degrees of freedom, i.e. phonons, which carry the EPC supposed to drive superconductivity. To this end we performed extensive measurements using high energy-resolution inelastic x-ray scattering (IXS) in Cu-intercalated as well as pressurized $1T$-TiSe$_2$ samples focusing on the evolution of the CDW soft phonon mode. In order to judge the impact on superconductivity, we measured at temperatures down to T = 5 K, only a few Kelvin above the reported superconducting transition temperatures for the respective intercalation and pressure levels.

We measured samples with $x = 0, 0.06$ and $0.09$, the latter two having superconducting transition temperatures of $T_c = 2.7$ K and $3.7$ K, respectively [37,38]. The CDW transition temperatures are $T_{CDW} \approx 200$ K, $50$ K and $< 6$ K for $x = 0, 0.06$ and $0.09$, respectively, in agreement with previous reports for similarly intercalated samples [3]. The latter were determined from measurements of the elastic intensity at $\boldsymbol{q}_{CDW}$ for $x = 0$ [16] and $0.06$ (Fig. 3a). We did not observe a corresponding increase of the elastic intensity in the $x = 0.09$ sample. However, measurements close to $\boldsymbol{q}_{CDW}$ reveal a constant softening of the mode energy (Fig. 2b) indicating the absence of a CDW phase transition temperature below which the energy of the soft mode is supposed to harden on further cooling [39]. IXS experiments were performed on the HERIX spectrometer at sector 3 of the Advanced Photon Source, Argonne National Laboratory. We investigated the dispersion along the $A - L$ line in reciprocal space at $\boldsymbol{Q}$ = (-h, 1, 0.5), $0 \leq h \leq 0.5$, where we have strong intensities of the CDW soft phonon mode [16]. Low temperature raw data taken in pure $1T$-TiSe$_2$ close to the $L$ point (Fig. 3c) reveal the lowest-energy phonon at $\omega \approx 11$ meV. In fact, the strong elastic tails of the superlattice peak at the $L$ point below $T_{CDW}$ prohibited IXS measurements in the ordered state at $\mathbf{q}_{CDW}$ and we therefore present data taken at $h = 0.45$. The phonon peak is well-defined and we see no indication of a line broadening on approaching the $L$ point, behavior which is only expected, and indeed observed in the vicinity of $T_{CDW} \approx 200$ K [16]. For the $x = 0.06$ and 0.09 samples we carried out IXS measurements in the temperature range $6\,K \leq T \leq 190$ K.

In Figure 3b,c we concentrate on the low temperature behavior of the soft phonon mode in intercalated samples. The data reveal very broad phonons close to zero energy. The observed dispersion along the $A - L$ line is different in the two intercalated samples (Fig. 3d): for $x = 0.06$ we observe a linear dispersion extrapolating to zero energy at the $L$ point, whereas for $x = 0.09$ the dispersion becomes very flat in the range $0.4 \leq h \leq 0.5$.

Regarding superconductivity, the contribution of a particular phonon mode $(\mathbf{q}, \vartheta)$ to the Eliashberg function (eq. 2) and, hence the superconducting $T_c$, is given by the line width divided by the phonon energy, $\gamma_{\mathbf{q}\vartheta}/\omega_{\mathbf{q}\vartheta}$. Whereas this ratio is very small for pure $1T$-TiSe$_2$ at low temperatures, we observe strongly increasing values on approaching the $L$ point for $x = 0.06$ and an even more pronounced enhancement for $x = 0.09$ (Fig. 3e). Hence, the low temperature properties of the CDW soft phonon mode are consistent with the DFPT model discussed above, in which the EPC of the soft mode is the main driving force for the superconducting phase near the CDW critical point. In particular, the results for $\gamma_{\mathbf{q}\vartheta}/\omega_{\mathbf{q}\vartheta}$ qualitatively agree with the increase of the superconducting transition temperature from $T_c(x = 0.06) = 2.7$ K to $T_c(x = 0.09) = 3.7$ K.

A comparison of the experimental results with the calculated values for $\gamma_{\mathbf{q}\vartheta}/\omega_{\mathbf{q}\vartheta}$ near the critical point, i.e. $\sigma = 0.15$ eV (solid line in Fig. 3e), shows that the observed values are even bigger than predicted. However, these discrepancies likely originate in a general underestimation of line widths of soft phonon modes by DFPT, which we reported previously at the CDW transition temperatures in $1T$-TiSe$_2$ [16] and NbSe$_2$ [32,40] as well as for low energy phonon modes in superconducting YNi$_2$B$_2$C [27]. Anharmonic contributions, not considered in DFPT, are likely to be present in the close vicinity of a structural phase transition. However, the qualitative momentum dependences of the observed linewidths in $1T$-TiSe$_2$ [16] as well as in other strong coupling materials such as $2H$-NbSe$_2$ [32,40] and superconducting YNi$_2$B$_2$C [27,41] are well described in DFPT. Hence, we argue that the observed linewidth is still reflecting the evolution of EPC in Cu$_x$TiSe$_2$. Furthermore, we recently showed that in $2H$-NbSe$_2$, the soft mode's linewidth at low temperatures is still dominated by EPC although DFPT is underestimating the experimentally observed value [42]. Summarizing, our results from IXS in Cu-intercalated $1T$-TiSe$_2$ are consistent with a model of solely phonon-mediated superconducting pairing near the critical point of the CDW phase.

## V. TiSe$_2$ under pressure

Using diamond anvil cells (DACs) to study emergent SC in TiSe$_2$ provides a powerful tool to investigate the competition between CDW order and SC in a stoichiometric sample. However, IXS at high pressures *and* low temperatures has an intrinsic low scattering rate because of the reduced sample volume compared to ambient pressure measurements. In our DAC experiments the linear dimension within the *a-b*-plane of the sample was restricted to 100 µm, with a corresponding thickness along the $c$ axis of $10 - 20$ µm ($70 - 100$ µm for ambient pressure experiments). We could partly compensate for the corresponding decrease in count rate by performing our experiment on the HERIX spectrometer located at the ID-30 beamline, which allowed us to measure in a higher Brillouin zone, i.e. $\mathbf{Q} = (-1 - h, 3, 0.5)$, as compared to our measurements in Cu$_x$TiSe$_2$ performed at ID-3. We measured at five different pressures and a total of seven points in the $(T, p)$ phase diagram. At the lower pressure values, $p = 1.75$ GPa, 2.5 GPa and 4.3 GPa, we investigated the soft mode behavior at the CDW transition temperature. The observed transition temperatures are in good agreement with a recent report from x-ray diffraction, which found that the critical pressure is as high as 5.1 GPa [5]. The dispersions along the A-L line at these $(T_{CDW}, p)$ points look very similar to the ambient pressure dispersion at $T = T_{CDW}$ (Ref. [16]) except for a pressure-induced hardening of all phonon energies (Fig. 4).

We now discuss the pressure dependence of the soft mode close to or at $\mathbf{q}_{CDW}$ at low temperatures. Increasing the pressure up to 2.5 GPa does not dramatically change the energy of the soft mode, which remains located above 10 meV (Fig. 5a). Closer to the critical pressure, we find a clearly reduced energy $E = (4.6 \pm 1.5)$ meV (Fig. 5b), which hardens as pressure is increased further past the critical pressure to $p = 6.3$ GPa (Figs. 5c). We are closest to the superconducting phase in our measurement at $p = 2.5$ GPa, where the reported $T_c$ is close to its maximum value of 1.8 K [4]. However, the values for $\gamma_{q\lambda}/\omega_{q\lambda}$ do not exceed 0.15, i.e. they are similar to what we find in pure $1T$-TiSe$_2$ at ambient pressure (Fig. 3e). The model with a strong increase of the Eliashberg function due to an increasing EPC near the critical point of the CDW phase cannot explain the emergence of superconductivity at this pressure. In fact, $\gamma_{q\lambda}/\omega_{q\lambda}$ increases as pressure is raised to $p = 4.3$ GPa (Fig. 5e), while the superconducting transition temperature is lowered and disappears for $p > 3.8$ GPa [4]. Going beyond the reported critical pressure of $p_c = 5.2$ GPa (Ref. [5]), we find a still slightly enhanced $\frac{\gamma_{q\lambda}}{\omega_{q\lambda}} = 0.3$ at $p = 6.3$ GPa, which is further reduced to the ambient pressure value of around 0.1 at $p = 9.6$ GPa (Fig. 5e). Hence, we conclude that the emergence of a pressure-induced SC dome in *1T*-TiSe$_2$ is unconventional, in the sense that it is inconsistent with a scenario in which the superconducting pairing is controlled by the EPC associated with a purely phononic soft mode. This conclusion is corroborated by the observation that no SC is reported near the critical pressure down to $T = 0.1$ K,[4] whereas a maximal superconducting $T_c$ would be expected if both CDW and superconductivity were promoted solely by EPC (see Fig. 1 and discussion above).

Our observations can however be explained in a scenario in which the soft mode is a hybrid phonon-exciton mode and the effective pairing interaction giving rise to superconductivity is strongly dependent upon the degree of hybridization between exciton and phonon. The presence and importance of excitons on the CDW order has been extensively discussed for the undoped compound *1T*-TiSe$_2$ at ambient pressure [11,14-16,43-47]. Below, we demonstrate theoretically that the pressure dependence of hybridazation between exciton and phonon modes can explain our experimental observations throughout the full $(T, p)$ phase diagram.

## VI. Hybrid phonon-exciton model

The band structure of *1T*-TiSe$_2$ is well-known to be close to a zero-gap indirect semiconductor.[48,49] This feature, along with the associated low electronic screening, has led to the suggestion that this is an ideal material for the realization of the elusive excitonic insulator state.[50,51] This state is a CDW whose formation is mediated not by phonons, but instead by excitons, or bound particle-hole pairs. Although it has been suggested several times that some experimentally observed features of *1T*-TiSe$_2$ are consistent with its CDW being an excitonic insulator,[14,52] it has similarly been claimed repeatedly that all of its properties can be explained within a purely phonon-based scenario.[15,47] More recently, it was realized that if excitons exist within *1T*-TiSe$_2$, they must necessarily couple strongly to the omnipresent phonons, resulting in the emergence of hybrid phonon-exciton modes.[45,46] The hybridization is unavoidable, because the coupling between excitons and phonons is given by the same EPC matrix element $g_{k+qv',kv}^{q\vartheta}$ which appears in equation (1), and the CDW in practice must therefore always be mediated by a mode that is partly exciton and partly phonon. Consistent with this observation, signatures of contributions to the CDW formation arising from both phonon-mediated and Coulomb interactions have recently been identified in time-resolved experiments.[43]

Intercalation with Cu atoms shifts the chemical potential of *1T*-TiSe$_2$ by 0.1 eV.[53,54] Since the circumstances in the pristine material are nearly ideal for the formation of excitons, they deteriorate quickly upon intercalation, leaving a dominant role for the phonons in the CDW formation in this part of the phase diagram. As explained above, such a scenario naturally leads to a superconducting dome centered around the critical value of intercalation.

In contrast, the application of pressure leaves the band structure and chemical potential relatively unaffected, and we argue here that in this case, the hybridization between phonon and exciton modes provides a natural way for the superconducting dome to move away from the critical pressure. To illustrate this, we first consider the theoretical situation of a purely phonon-mediated CDW, accompanied by a massive exciton that does not interact with the phonon modes at all. Starting from the normal state at zero temperature and high pressure, and going towards the ambient pressure CDW phase, the phonon mode at the $L$ point will initially soften as the critical pressure $p^*$ of the phonon-mediated CDW is approached (dashed line in Fig. 6a). The excitonic energy on the other hand, being determined solely by the Coulomb interaction, may remain constant in energy (dash-dotted line in Fig. 6a). Decreasing pressure further within the CDW phase below $p^*$, the soft phonon mode becomes an amplitude mode for the CDW lattice displacements, which hardens as the order develops (dashed line below $p^*$ in Fig. 6a). If we now consider interaction and hybridization between the exciton and phonon, there will be an avoided crossing between their dispersions, which results in the low energy hybrid mode reaching zero energy at a pressure $p_c > p^*$ (thick solid line in Fig. 6a). As pointed out above, the coupling between phonon and exciton is provided by the EPC matrix element and cannot be turned off in any real material, so that $p_c$ corresponds to the critical pressure at which the CDW is experimentally observed to form. The low temperature pressure dependences of the observed excitations for 2.5 GPa $\leq p \leq$ 10 GPa are shown in Figure 6b. The optic mode near 20 meV shows a monotonically increasing energy (squares), whereas the soft mode energies (dots/circles) suggest a minimum energy at $P \approx P_c$. We interpret this dispersing soft mode to be the low-energy hybrid phonon-exciton mode indicated by the solid black line in Figure 6a.

To model the influence of hybridization on the emergence of the superconducting dome, we start from the normal phase at high pressures, and employ a one-dimensional, two-band electronic model:

$$\widehat{H}_{\text{electron}} = \sum_k E_v(k)\hat{v}_k^\dagger \hat{v}_k + E_c(k)\hat{c}_k^\dagger \hat{c}_k \qquad (4)$$

Here $\hat{v}_k^\dagger$ and $\hat{c}_k^\dagger$ create electrons in the valence and conduction bands respectively. Because the low-energy physics is dominated by the top of the valence band and the bottom of the conduction band, we employ the simplest possible band structure with quadratic maxima and minima, $E_{c,v}(k) = E_{c,v}^0 + 2t_{c,v}\cos(ka)$. The hopping integrals $t_{c,v}$ are chosen such that the band curvatures at the extrema are consistent with the experimentally observed and numerically estimated curvatures in the literature in the normal phase at ambient pressure.[48,55] The energies $E_{c,v}^0$ are chosen to yield a small indirect band gap.[48,49]

We then include phonon and exciton modes with momentum $Q = \Gamma L$, which is the momentum connecting the top of the valence band to the bottom of the conduction band:

$$\widehat{H}_{\text{boson}} = \epsilon_{\text{ph}}\hat{a}^\dagger \hat{a} + \epsilon_{\text{exc}}\hat{b}^\dagger \hat{b} + g(\hat{a}^\dagger \hat{b} + b^\dagger \hat{a}) \qquad (5)$$

Notice that $\epsilon_{\text{ph}}$ is the renormalized phonon energy, which already includes the effect of particle-hole fluctuations, and which would correspond to the experimentally observed mode if no excitons were present (the dashed line in Fig. 6a). The exciton energy (corresponding to the dash-dotted line in Fig. 6a) is determined by the combination of Coulomb-induced excitonic binding energy and the value of the indirect gap between the valence and conduction bands. The interaction between the two bosonic modes is provided by $g$, the EPC matrix element.

Finally, a superconducting pairing interaction can arise according to the usual BCS mechanism from the interaction between electrons and renormalized phonons:

$$\hat{H}_{\text{int}} = \sum_k \gamma (\hat{a}^\dagger + \hat{a})(\hat{v}_k^\dagger \hat{c}_{k+Q} + \hat{c}_{k+Q}^\dagger \hat{v}_k) \qquad (6)$$

The coupling to renormalized phonons $\gamma$ is proportional to the coupling to bare phonons $g$. Integrating out all bosonic modes leads to effective Cooper-pairing interactions, both within and between electronic bands. Following the standard BCS derivation, the effective pairing interaction in the presence of hybridized modes will be proportional to:

$$V_{\text{eff}} \propto |\alpha|^2 \frac{g^2}{\omega_-} + |\beta|^2 \frac{g^2}{\omega_+} \qquad (7)$$

Here the lowest energy hybrid mode with energy $\omega_-$ (the lowest solid line in Fig. 6a) is given by $|hybrid\rangle = \alpha |phonon\rangle + \beta |exciton\rangle$. Since the amount of hybridization depends on the difference in energy between the bare modes, $\alpha$ is pressure dependent. Together with the pressure dependence of $\omega_\pm$ this determines the evolution of the BCS pairing potential $V_{\text{eff}}$, and hence the superconducting $T_c$ under pressure.

To model this pressure dependence, we assume that in the absence of excitons, the renormalized phonon mode would soften to zero energy and induce a CDW transition at some pressure $P^*$, as indicated by the dashed line in Figure 6a. This is modeled by a heuristic, pressure-dependent analytic form of the renormalized phonon energy, which goes to zero at $P^*$ and extrapolates to the linearly increasing behavior expected for a bare phonon mode far from the CDW transition:

$$\epsilon_{\text{ph}}(P) = \epsilon_0 \left(1 + \eta \frac{P}{P^*}\right) \tanh\left(\frac{2}{\sigma}\left(\frac{P}{P^*} - 1\right)\right) \qquad (8)$$

The value of the slope $\eta$ is chosen to agree with the linear increase observed for high-energy phonon modes in TiSe, shown in Figure 6b, which are believed to be unaffected by the CDW formation. The width $\sigma$ of the region over which the renormalized phonon softens, is used as a free parameter determining the width of the superconducting dome. The value of the hybridization parameter $g$ finally, is chosen such that the lowest hybrid mode softens to zero energy at the experimentally observed onset of CDW order, $P_c > P^*$, as seen in Figure 6a.

With all of these parameters fixed, the BCS superconducting transition temperature can be calculated as a function of pressure. The combination of a sharp decrease of $\alpha$ at $p > p^*$, and the peaked structure of $1/\omega_-$ around $p_c$ result in a superconducting dome centered between $p^*$ and $p_c$, as shown in Figure 6c. Here we chose parameter values in a range which is believed to be reasonable for *1T*-TiSe$_2$, and which reproduce the experimental observation of the superconducting dome being contained entirely within the CDW phase.

The combined influence of intercalation and pressure is shown in the full phase diagram of Figure 6d. Here, the CDW formation is always caused by a hybrid mode, whose exciton content increases strongly as a function of pressure beyond the crossover scale $p^*$, while it uniformly decreases as the level of intercalation rises. The superconducting phase is always BCS-like, originating only from the electron-phonon coupling, which allows its position with respect to the edge of the CDW phase to be used as a measure of the exciton content in the hybrid mode underlying CDW formation.

## VII. Discussion

The ability to control the relative exciton and phonon content of the soft mode in the CDW phase of TiSe$_2$ using the tuning parameters of pressure and intercalation, as suggested by the combination of experimental, numerical and theoretical results presented here, has both practical and fundamental implications. For example, because the soft mode at $p_c$ has a strongly increased excitonic character, the CDW phase near the critical point should be much closer to being an excitonic insulator than its counterpart at ambient pressure. If the excitonic content could be increased even further (using an alternative tuning parameter), the CDW state could become practically indistinguishable from this long-sought but elusive state of matter. Suppressing the CDW phase at that point would yield a quantum critical point at which the critical fluctuations are almost entirely excitonic in nature. While this could be thought of as the ideal breeding ground for the equally long-sought exciton-mediated superconducting phase [56-58], we note that no superconducting $T_c$ was detected down to 0.1 K [4].

The fact that we can use a combination of observed soft mode energies and line widths, DFPT calculations, and a simple form of BCS theory as a direct indication of the varying exciton content throughout the phase diagram of TiSe$_2$, provides a novel characterization tool in the search for excitonic phases of matter in other materials, as well as a direct test of the novel physics of order mediated by hybrid soft modes.

## VIII. Summary

We have investigated the role of the CDW soft mode regarding the emerging superconducting domes in *1T*-TiSe$_2$ via *ab-initio* lattice dynamical calculations and inelastic x-ray scattering. While we find good agreement between the purely phonon-driven scenario of DFPT and our experimental results for $Cu_xTiSe_2$, *1T*-TiSe$_2$ under pressure behaves qualitatively different. We determine a central role of the hybridization of excitons with phonons in positioning the superconducting dome and provide a model for the full $(T, x, p)$ phase diagram of *1T*-TiSe$_2$. These results show that *1T*-TiSe$_2$ under pressure is close to the elusive phase of an excitonic insulator.

.

**Acknowledgements:**

M.M. and F.W. were supported by the Helmholtz Society under contract VH-NG-840. S.Z. and J.v.W. acknowledge support from a VIDI grant financed by the Netherlands Organisation for Scientific Research (NWO). S.R. was supported by the US Department of Energy, Office of Science, Materials Science and Engineering Division. Research conducted at ANLs Advanced Photon Source was sponsored by the Scientific User Facilities Division, Office of Basic Energy Sciences, U.S. Department of Energy.

## Appendix A: Structural distortion for $\sigma = 0.1$ eV

The lattice parameters in all calculations were fixed to those observed in experiment.[16,34] The reason is that the $c/a$ ratio was found to be very sensitive to the used functionals (GGA [general gradient approximation] and LDA [local density approximation]) in agreement with recent theoretical work.[34] Whereas GGA calculates $\left(\frac{c}{a}\right)_{GGA} = 1.924$ ($a_{GGA} = 3.5365$ Å), LDA predicts $\left(\frac{c}{a}\right)_{LDA} = 1.687$ ($a_{LDA} = 3.4412$ Å), which is closer to the experimental value $\left(\frac{c}{a}\right)_{exp} = 1.697$ ($a_{exp} = 3.54$ Å). However, the absolute values of the lattice constants on their own are significantly underestimated in LDA. The optimized calculations vary in lattice dynamical calculations in that LDA predicts only a modest phonon anomaly with finite energy at the $L$ point whereas GGA calculations yield large negative values for the square of the lowest phonon energy, i.e. imaginary energies, at the $L$ point. It has been shown in Ref. [34] that these functionals apparently have a problem in describing the bonding between TiSe$_2$ layers via van-der-Waals forces. Using the experimental lattice constant and only relaxing internal coordinates results in lattice dynamical calculations and structural distortions independent of the functional used,[34] and this is what we used for all presented calculations.

Calculations performed in the trigonal unit cell ($P\bar{3}m1$, #164) and small smearing parameters $\sigma = 0.1$ eV and 0.124 eV demonstrate that the lattice is unstable towards a structural distortion involving the doubling of the unit cell in all three directions. Therefore, lattice dynamical calculations were done in a $2 \times 2 \times 2$ supercell. The values for the lattice parameters of the supercell were fixed to two times the experimental lattice constants of the small trigonal cell [16,34] (Table 1). The relaxed atomic positions within the supercell are given in relative units of the supercell lattice parameters in Table 1.

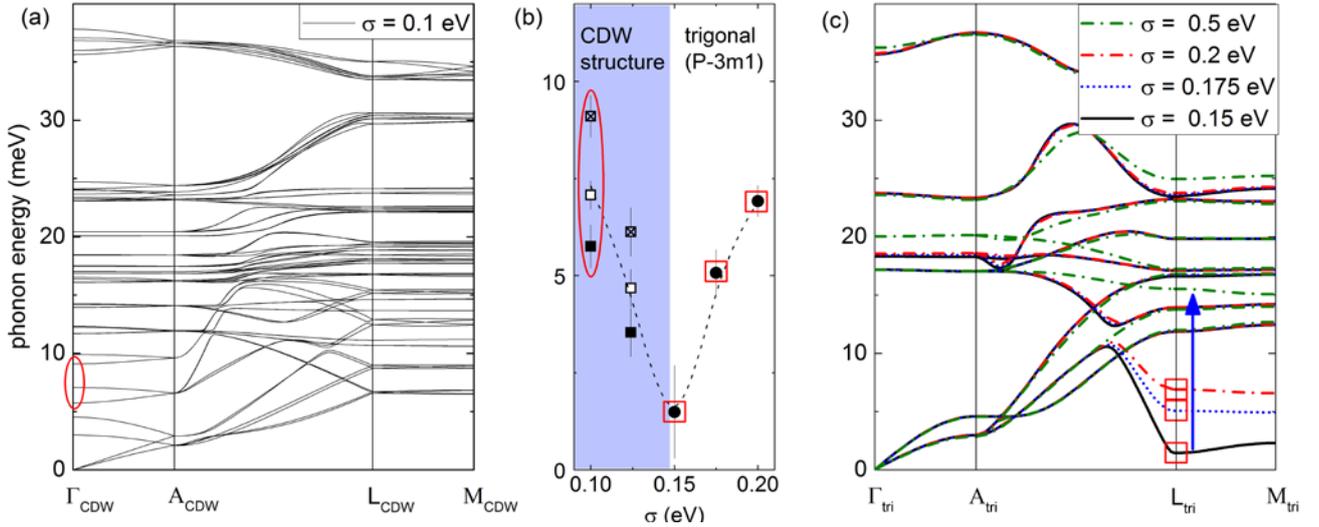

**Figure 1:** *(c)* Phonon dispersion lines along various high symmetry directions of *1T*-TiSe$_2$ for $\sigma = 0.1$ eV calculated in the $2 \times 2 \times 2$ supercell corresponding to the distorted structure in the CDW phase. The red ellipsoid marks the soft modes carrying strong EPC folded back from the three $L_{tri}$ points of the trigonal high temperature structure. *(b)* Energies of CDW soft mode(s) for $\sigma = 0.1 - 0.2$ eV. Calculations for $\sigma \leq 0.124$ eV were performed in the $2 \times 2 \times 2$ supercell corresponding to the structure in the CDW phase. Here, we plot the energies of the three $\Gamma_{CDW}$ modes carrying the strong EPC [see (a)]. Error bars correspond to the calculated values of the electronic contribution to the phonon linewidth $\gamma$. The line is a guide to the eye. Squares and ellipsoid mark the soft mode energies denoted similarly in panels (a) and (c). *(c)* Phonon dispersion lines along various high symmetry directions of *1T*-TiSe$_2$ for different smearing parameters $\sigma = 0.15 - 0.5$ eV in the trigonal structure. The vertical (blue) arrow marks the hardening of the CDW soft mode on going from $\sigma = 0.15$ eV to $0.5$ eV. Red squares denote the phonon energies of the soft mode also shown in (b).

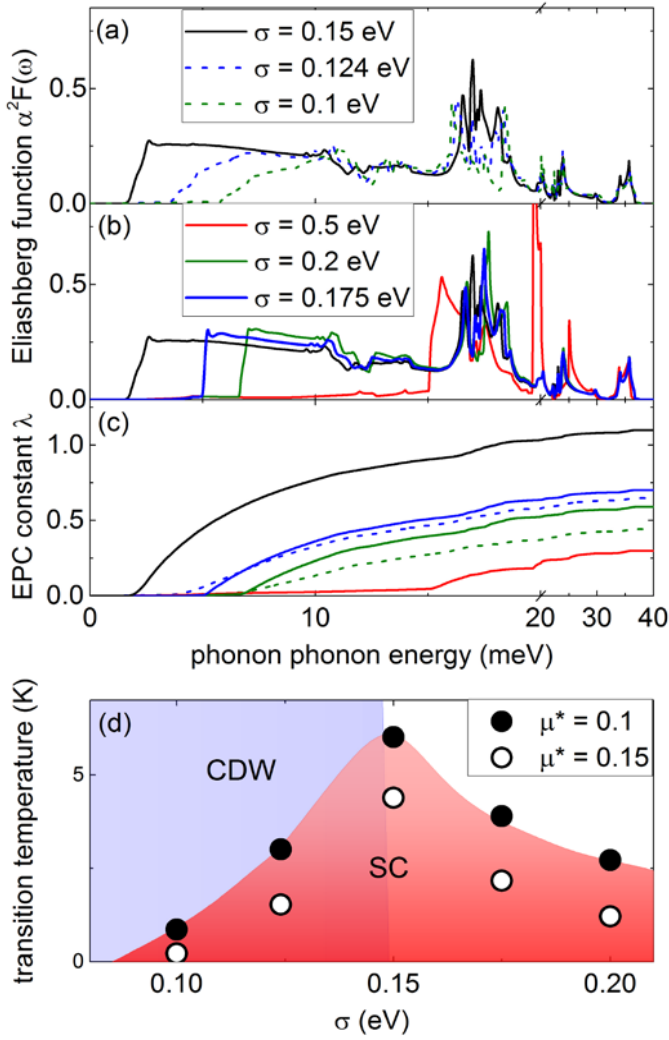

**Figure 2:** *(a,b)* Calculated Eliashberg function $\alpha^2 F(\omega)$ and *(c)* electron-phonon-coupling constant $\lambda(\omega)$ of *1T*-TiSe$_2$ for different smearing parameters $\sigma = 0.1 - 0.5$ eV from DFPT. *(d)* Superconducting transition temperatures $T_c$ for *1T*-TiSe$_2$ (dots,circles) as function of the electronic smearing $\sigma$ estimated from results shown in (a)-(c) using effective electron-electron interactions of $\mu^* = 0.1$ (dots) and 0.15 (circles) (see text). The CDW distorted phase is stable for $\sigma < 0.15$ eV (blue shaded area). Lines are guides to the eye.

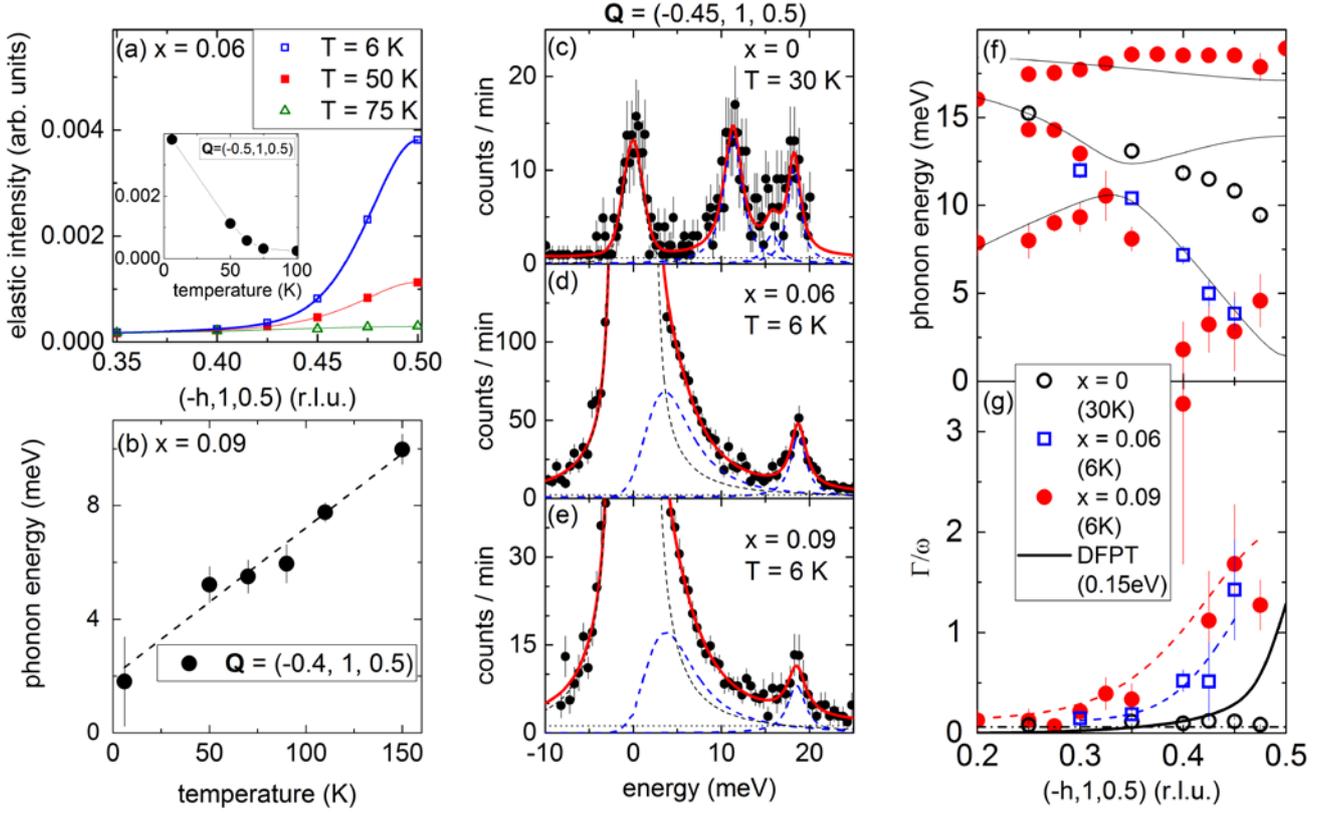

**Figure 3: Low temperature IXS in Cu$_x$TiSe$_2$:** *(a)* Elastic intensity measured in the $x = 0.06$ sample close to the CDW ordering wavevector $\mathbf{q}_{CDW}$ at three different temperatures. This inset shows the temperature dependence of the intensity at $\mathbf{q}_{CDW}$. *(b)* Observed energies of the soft mode close to $\mathbf{q}_{CDW}$ in the $x = 0.09$ sample as function of temperature. *(c),(d),(e)* Raw IXS data taken close to the CDW superlattice peak position at low temperatures in intercalated samples with $x = 0, 0.06,$ and $0.09$, respectively. Solid (red) lines are fits consisting of DHO functions for the phonons (thick dashed lines), a pseudo-voigt function for the resolution limited elastic line (thin dashed line) and the estimated experimental background (dotted straight line). Note that the comparably strong elastic lines in *(d),(e)* result from increased incoherent scattering in the intercalated, i.e., intrinsically dirty samples and not from CDW order. *(f)* Observed dispersions for different Cu concentrations $x$ at low temperatures. For clarity, we show all observed phonon modes only for x = 0.09 (dots) and only the soft mode energies for $x = 0$ (open circles) and 0.06 (open squares). Solid lines are DFPT calculations with $\sigma = 0.15$ eV. *(g)* Ratio of the soft mode's line width $\Gamma$ and its energy $\omega$ along the A-L at low temperatures. The thick line is the result of DFPT with $\sigma = 0.15$ eV. Dashed lines are guides to the eye.

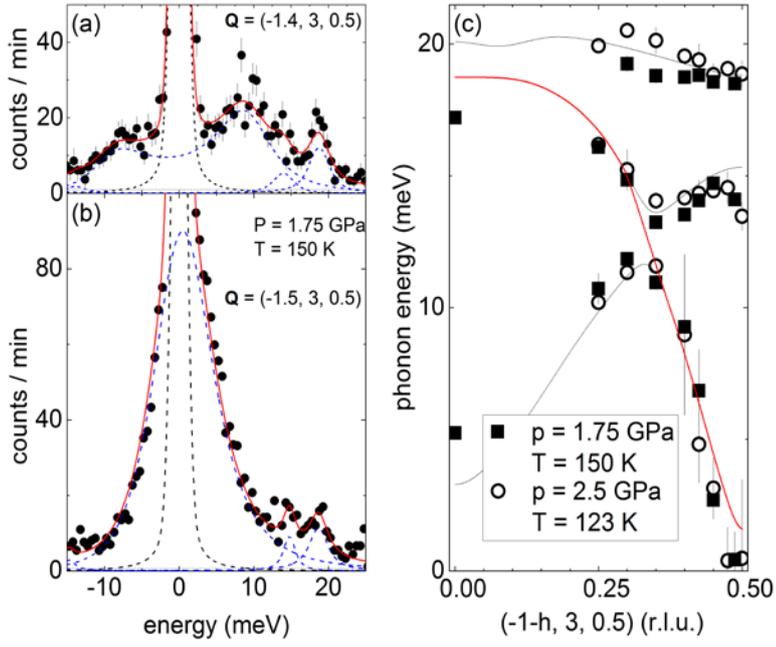

**Figure 4:** *(a,b)* Raw IXS data taken *(a)* close to and *(b)* at the CDW superlattice peak position for $T_{CDW}(p = 1.75\ \text{GPa}) = 150\ \text{K}$ (same symbol/line code as in Figs. 3a-c). *(c)* Dispersion of the soft phonon mode for two sets of $(T_{CDW}, p)$ values. Thin lines denote DFPT calculations with $\sigma = 0.15\ \text{eV}$ but scaled with a factor of 1.1. The thick (red) line marks the dispersion of the soft mode character across the anti-crossing at $h \approx 0.3 - 0.35$.

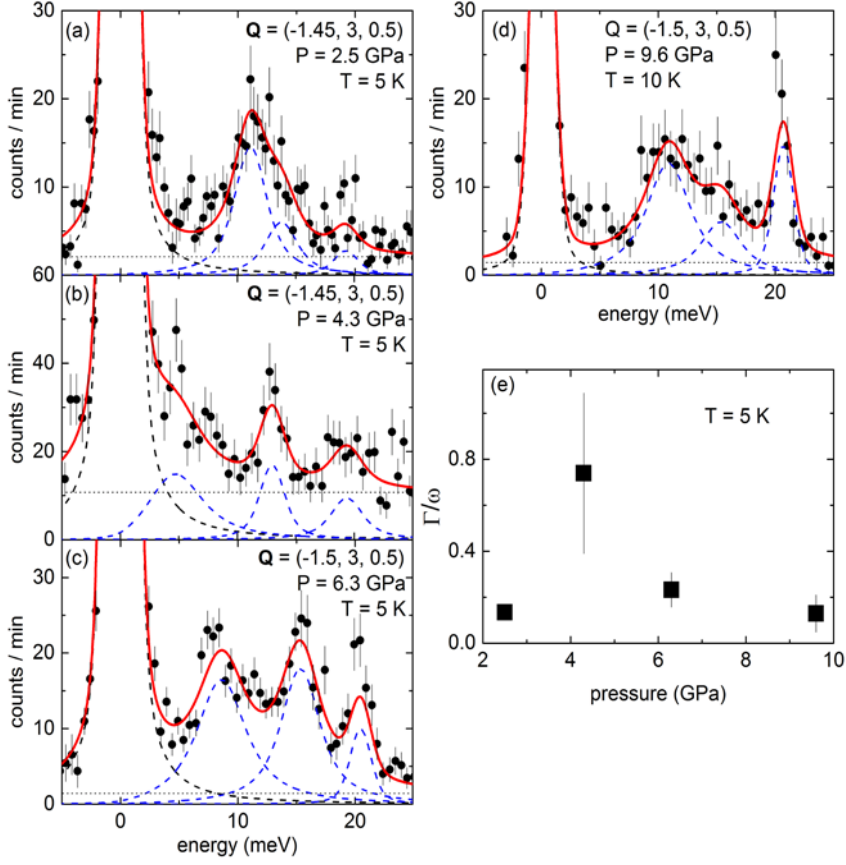

**Figure 5:** *(a)-(d)* Raw IXS data taken at $T = 5$ K and increasing pressures 2.5 GPa $\leq p \leq$ 9.6 GPa (same symbol/line code as in Figs. 3a-c). Data at $p < p_c = 5.1$ GPa were taken slightly away from $\mathbf{q}_{CDW}$. At higher pressures IXS scans were performed at $\mathbf{q}_{CDW}$. *(e)* Ratio of the soft mode's line width $\Gamma$ and its energy $\omega$ as shown in the fits in (a)-(d).

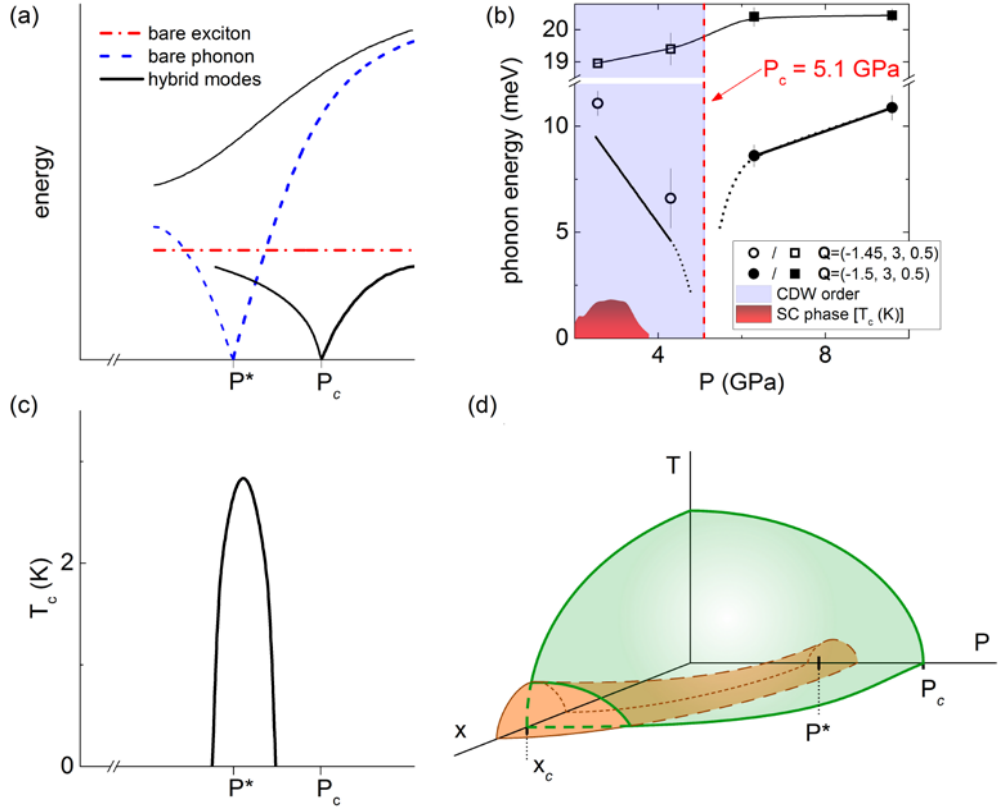

**Figure 6:** *(a)* Sketch of the excitation energies appearing in the theoretical model without (dashed lines) and including mode hybridization (solid lines) (see text). *(b)* Summary of the low temperature excitations observed in IXS (see text). Solid lines are guides to the eye of the measured energies including an estimate for the actual soft mode energy at the CDW ordering wavevector for $P < P_c$ based on the observed difference of 1.5 meV between IXS at $\boldsymbol{Q} = (-1.45, 3, 0.5)$ and energies observed in Raman scattering [59]. Dotted lines indicate the expected behavior of the soft mode approaching a quantum critical point at $P = P_c$. *(c)* The superconducting transition temperature as a function of pressure according to equation (7). This figure uses the parameter values $t_v = 2t_c = 3.0$ meV, $E_c^0 - E_v^0 = 20$ meV, $\varepsilon_0 = 17$ meV, $E_{exc} = 5.0$ meV, $g = 10$ meV, and $\gamma = 17.8$ meV, as well as $\eta = 0.18$, $\sigma = 0.75$, and $p^*/p_c = 0.6$. *(d)* The phase diagram emerging from the theoretical model (see text).

|          | Lattice constants |          |          |          | $a = b = 7.07$ Å |          | $c = 12.01$ Å |          |
|----------|---------|----------|----------|----------|----------|----------|----------|----------|
|          | Ti(1)   | Ti(2)    | Ti(3)    | Ti(4)    | Ti(5)    | Ti(6)    | Ti(7)    | Ti(8)    |
| x (l.u.) | 0       | 0.49526  | -0.49526 | 0        | 0        | 0.50474  | -0.50474 | 0        |
| y (l.u.) | 0       | 0        | -0.49526 | 0.49526  | 0        | 0        | -0.50474 | 0.50474  |
| z (l.u.) | 0.25    | 0.25     | 0.25     | 0.25     | 0.75     | 0.75     | 0.75     | 0.75     |
|          | Se(1)   | Se(2)    | Se(3)    | Se(4)    | Se(5)    | Se(6)    | Se(7)    | Se(8)    |
| x (l.u.) | 0.33323 | -0.16533 | -0.1679  | -0.66667 | -0.33323 | 0.1679   | 0.16533  | 0.66667  |
| y (l.u.) | 0.16533 | 0.1679   | -0.33323 | -0.33333 | -0.1679  | -0.16533 | 0.33323  | 0.33333  |
| z (l.u.) | 0.37557 | 0.37557  | 0.37557  | 0.37551  | 0.12443  | 0.12443  | 0.12443  | 0.12449  |
|          | Se(9)   | Se(10)   | Se(11)   | Se(12)   | Se(13)   | Se(14)   | Se(15)   | Se(16)   |
| x (l.u.) | 0.33323 | -0.1679  | -0.16533 | -0.66667 | -0.33323 | 0.16533  | 0.1679   | 0.66667  |
| y (l.u.) | 0.1679  | 0.16533  | -0.33323 | -0.33333 | -0.16533 | -0.1679  | 0.33323  | 0.33333  |
| z (l.u.) | 0.87557 | 0.87557  | 0.87557  | 0.87551  | 0.62443  | 0.62443  | 0.62443  | 0.62449  |

**Table 1:** Lattice constants and atomic positions for the $2 \times 2 \times 2$ supercell calculations with $\sigma = 0.1$ eV. Atomic positions $(x, y, z)$ are given in units of the lattice vectors (lattice units, l.u.), the length of which were fixed to twice the values of the experimental lattice constants of the trigonal high temperature structure.